\documentclass[aps,prd,preprint,groupedaddress,showpacs,nofootinbib]{revtex4}
\usepackage{graphicx,float}
\usepackage{amsfonts}
\usepackage{subfig}
\usepackage{caption}
\usepackage{color}
\begin{document}


\title{A no-hair theorem for spherically symmetric black holes in $R^2$ gravity}


\author{Joseph Sultana}
\email{joseph.sultana@um.edu.mt}
\affiliation{Department of Mathematics, Faculty of Science,
University of Malta, Msida, Malta}
\author{Demosthenes Kazanas}
\email{Demos.Kazanas-1@nasa.gov}
\affiliation{Astrophysics Science Division, NASA/Goddard Space Flight Center, Greenbelt, Maryland 20771, USA}

\date{\today}

\begin{abstract}

\baselineskip=14pt

\parskip = 2mm

In a recent paper Ca\~{n}ate (CQG, {\bf 35}, 025018 (2018)) proved a no hair theorem to static and spherically symmetric or stationary axisymmetric black holes in general $f(R)$ gravity. The theorem applies for isolated asymptotically flat or asymptotically de Sitter black holes and also in the case when vacuum is replaced by a minimally coupled source having a traceless energy momentum tensor. This theorem excludes the case of pure quadratic gravity,
$f(R) = R^2$. In this paper we use the scalar tensor representation of general $f(R)$ theory to show that there are no hairy black hole in pure $R^2$ gravity. The result is limited to spherically symmetric black holes but does not assume asymptotic flatness or de-Sitter asymptotics as in most of the no-hair theorems encountered in the literature. We include an example of a static and spherically symmetric black hole in $R^2$ gravity with a conformally coupled scalar field having a Higgs-type quartic potential.

\end{abstract}

\pacs{04.50.Kd, 04.70.Bw, 04.20.Jb}
\keywords{no-hair theorems, scalar field, black holes, $R^2$ gravity}

\maketitle

\section{Introduction}

\baselineskip=15pt

Following the \textit{no-hair conjecture} (also known as the no-hair theorem (NHT))\cite{israel67,carter71,wheeler71} which precludes the existence of asymptotically flat black holes solutions of Einstein's field equations with additional parameters (the so called hair) besides the mass, angular momentum and charge of the black hole, there have been several generalizations of this conjecture, particularly in other theories of gravity such as Brans-Dicke theory, scalar tensor theories \cite{bekenstein72,hawking72,johnson72,zannias95,bekenstein95,bekenstein96,saa96,bronnikov01,nelson10,sotiriou12,bhatta15} and Horndeski gravity \cite{herdeiro15,sotiriou15}. Eventually a number of counterexamples to this conjecture consisting of black holes with scalar hair have been obtained in other theories such as Einstein-Yang-Mills Dilaton theory, Einstein-Dilaton-Gauss-Bonnet theory \cite{kanti96} and shift-symmetric Galileon theories \cite{sotiriou14}.

The discovery of the present accelerating cosmic expansion with type Ia supernovae led to an increased interest in modified theories of gravity. Instead of using dark energy as the source of the cosmic expansion these theories modify gravity to be repulsive on cosmological scales. A popular class of modified gravity are the $f(R)$ theories, in which the Lagrangian in the Einstein-Hilbert action is replaced by an arbitrary function of the Ricci scalar $R$. So far there are several specific $f(R)$ theories that have been suggested as producing a late time accelerated cosmic expansion. However only a few of these satisfy the classical tests of General Relativity. A number of authors \cite{canate15,bergliaffa11,dombriz09,gosh13,larranga11,psaltis08} have investigated various black hole vacuum solutions in $f(R)$ gravity with different asymptotics, and all of these were found to have a constant Ricci scalar curvature $R$. It will be shown later that when $R$ is constant the field equations of $f(R)$ gravity reduce to Einstein's field equations with an effective gravitation and cosmological constants. In this sense these solutions are trivial because they are equivalent to the black hole solutions in Eisntein's gravity. On the other hand in the presence of matter it is possible to find static and spherically symmetric \cite{multamaki08,kain07,hent08,babichev09,babichev10,upadhye09,yazadjiev14,miranda09} solutions having non-constant Ricci curvature. However these solutions are obtained numerically and it is not clear whether these can be matched to exact exterior solutions as in GR for example.

Recently Ca\~{n}ate \citep{canate17} presented a NHT for static and spherically symmetric black holes in $f(R)$ theory, which proves that black holes solutions in this theory are trivial (meaning that they have constant $R$). This result was also generalized to stationary and axisymmetric black holes. The theorem is valid for asymptotically flat or asymptotically de-Sitter spacetimes in vacuum or in the case when there is a minimally coupled matter with a traceless energy-momentum tensor. The proof assumes that the functions \footnote[1]{Note that the subscript in $f_R$ indicates differentiation of $f$ with respect to the Ricci scalar $R$} $U'(f_R) = (2f - Rf_R)/3$ and its derivative $U''(f_R)$ are positive definite, and therefore it excludes the case $f(R) \propto R^2$ in which case $U'(f_R) = U''(f_R) = 0$. This result is similar to an earlier no-hair theorem (see
Refs.\cite{bekenstein95,canate15,sudarsky95})
for static asymptotically flat black holes in $f(R)$ theory. In this case the authors have used the scalar tensor representation of $f(R)$ theory, and the theorem is actually proved in the Einstein conformal frame. Roughly it states that in an Einstein-$\phi$ system, in which the potential associated with the scalar field $\phi$ satisfies $U(\phi)\geq0$, any asymptotically flat static and spherically symmetric black hole is the hairless Schwarzschild solution. Later Sotiriou and Faraoni \cite{sotiriou12} extended this to a much more generalized class of scalar tensor theories, assuming stationarity instead of staticity but keeping asymptotic flatness and adopting the weaker convexity requirement that $U''(\phi) \geq 0$.


$R^2$-theory is the simplest example of the more general quadratic gravity theories whose action contains terms that are second order in the curvature tensor, namely
\begin{equation}
S = \int_{M}d^4x \sqrt{-g}(c_1C_{\mu\nu\rho\sigma}C^{\mu\nu\rho\sigma} + c_2R^2 + c_3\tilde{R}_{\mu\nu}\tilde{R}^{\mu\nu}),
\end{equation}
where $c_i$ are dimensionless coupling constants, $C_{\mu\nu\rho\sigma}$ is the Weyl tensor and $\tilde{R}_{\mu\nu} = R_{\mu\nu} - \frac{1}{4}g_{\mu\nu}R$ is the traceless part of the Ricci tensor $R_{\mu\nu}$. Partly due to their scale invariant properties which forbids the presence of any length scale, these theories have always generated a lot of interest.
Starting with Starobinsky's inflationary scenario \cite{starobinsky79,starobinsky80} of the early Universe in quantum-corrected general relativity with $f(R) = R + \alpha R^2$, we see that pure $R^2$ gravity is a very good approximation when the curvature and $R$ are large. Recently it was shown \cite{rinaldi14} that the inflationary phase of the early Universe is best described by an $R^2$ theory in favor of Starobinsky's model.
It is believed  that quadratic gravity is in general renormalizable and asymptotically free \cite{stelle77,adler82,fradkin81,avramidi85}. Yet they suffer from ghosts, except the simple case of $R^2$-gravity which is the only scale invariant quadratic theory that is ghost-free \cite{kounnas15}. The scale-invariance in quadratic gravity is extremely sensitive to external perturbations and so even the coupling with a small mass particle will break the scale symmetry. One can argue that for low curvatures, pure $R^2$ gravity is pathological because it lacks a Newtonian limit \cite{pechlaner66} and so astrophysical black holes are not relevant in this case. Yet however in the early inflationary phase primordial black holes could still be relevant. Static and spherically symmetric black hole solutions have been studied in $R^2$-gravity \cite{kehagias15_R2} and it was shown that Birkhoff's theorem does not apply in this case, so that besides the Schwarzschild solution, the theory admits a rich structure of vacuum solutions. Moreover rotating and topological black holes in $R^2$-gravity and their associated thermodynamics have also been studied \cite{cognola_15_2_R2, cognola15_R2}.

In this paper we present a NHT for black holes in $R^2$-gravity and therefore this extends the no-hair result obtained by Ca\~{n}ate \cite{canate17}.
The theorem is limited to static and spherically symmetric black holes, in vacuum or in the presence of conformally invariant matter. However unlike most
of the NHTs (including that in Ref. \cite{canate17}) no assumption is made about the black hole asymptotics. The proof of this theorem makes use of the
equivalence between $f(R)$ theories and Brans-Dicke (BD) theory \cite{olmo06} together with a recent \cite{faraoni17} NHT for spherical black holes in
scalar tensor gravity with vanishing scalar potential. So in the next section we give a brief introduction to $f(R)$ gravity and its equivalence with BD
theory with vanishing BD coupling constant. In section III we take the case $f(R)=R^2$ and we prove the NHT for static and spherically symmetric black
holes using the Jordan frame of the associated BD-theory. Then in section IV we present an example of a non-vacuum static and spherically symmetric black
hole in $R^2$-gravity with a conformally coupled massless scalar field having a Higgs-type quartic potential. The results are summarized and discussed in
the conclusion.

\section{Scalar tensor representation of f(R) gravity}

The action that corresponds to $f(R)$ gravity has the generic form
\begin{equation}
S = \frac{1}{2\kappa}\int\mathrm{d}^4x\sqrt{-g}f(R) + S^{(m)}, \label{fRaction}
\end{equation}
where $S^{(m)}$ is the matter part of the action and $\kappa$ is a constant. Varying this action with respect to the (inverse) metric $g^{\mu\nu}$ yields the following field equations
\begin{equation}
f_R(R)R_{\mu\nu} - \frac{f(R)}{2}g_{\mu\nu} = \nabla_{\mu}\nabla_{\nu}f_R(R) - g_{\mu\nu}\Box f_R(R) + \kappa T_{\mu\nu}, \label{fReqs}
\end{equation}
where $T_{\mu\nu}$ is the energy momentum tensor corresponding to the matter part of the action $S^{(m)}$. The trace of these equations then yields
\begin{equation}
\Box R = \frac{1}{3f_{RR}(R)}(\kappa T - 3f_{RRR}(R)(\nabla R)^2 + 2 f(R) - R f_{R}(R)), \label{trace}
\end{equation}
where $T = -\rho + 3p$ is the trace of $T_{\mu\nu}$. When used with (\ref{fReqs}) the above trace equation enables the field equations to take the Einstein form \begin{equation}
G_{\mu\nu} = \frac{\kappa}{f_R(R)}(T_{\mu\nu} + T_{\mu\nu}^{\mathrm{eff}}), \label{einstein-form}
\end{equation}
where
\begin{equation}
T_{\mu\nu}^{\mathrm{eff}} = \frac{1}{\kappa}\left[\frac{f(R) - Rf_R(R)}{2}g_{\mu\nu} + \nabla_{\mu}\nabla_{\nu}f_R(R)
- g_{\mu\nu}\Box f_R(R)\right] \label{emcfluid}
\end{equation}
is the effective source term due to the higher order curvature corrections. This is sometimes
described as the ``curvature fluid''. Using the Einstein form of the
field equations in (\ref{einstein-form}), one can define the effective gravitational
coupling $G_{\mathrm{eff}} = G/f_R(R)$. Hence for a positive $G_{\mathrm{eff}}$, we require that $f_R(R) > 0$.
Moreover stability of the theory would also require \cite{dolgov03,faraoni06}
$f_{RR}(R) >0$ for $R\geq R_0 > 0$ where $R_0$ is the value of the Ricci scalar today (calculated for the FLRW metric). The Einstein form of the field equations in (\ref{einstein-form})
implies also that besides the conservation of the total energy momentum, the
energy momentum tensor for the matter distribution $T_{\mu\nu}$ is independently conserved, i.e. $\nabla_\mu T^{\mu\nu}=0$. Note that in the case of a constant Ricci scalar $R=R_0$, Eqs. (\ref{einstein-form}) and (\ref{emcfluid}) reduce to Einstein's field equations with a redefined gravitational constant $G_{eff} = \frac{G}{f_{R}(R_0)}$ and an effective cosmological constant $\Lambda = \frac{1}{2}(R_0 - \frac{f(R_0)}{f_{R}(R_0)})$.

$f(R)$ theories can account for the late time acceleration, as well as the early inflationary period without the need of a cosmological constant $\Lambda$ or an extra scalar field as in scalar tensor theory (for a detailed review see Refs. \cite{sotiriou10,jaime12} and references therein). In this way dark energy effects arise solely from the geometry itself. However for any such $f(R)$ model to be a viable theory of gravitation, it should also agree with observations on all gravitational scales, including the Solar System tests. So for example the first proposed $f(R)$ model which accounts for the present day cosmic acceleration \cite{carroll04}, has the form $f(R) = R - \mu^{4}/R$, with a mass scale $\mu \sim H_0 \approx 10^{-33}\mathrm{eV}$. This however was soon ruled out due to a catastrophic instability \cite{dolgov03,nojiri03} and the violation of post-Newtonian tests of GR \cite{chiba03}. Despite this the $f(R)$ proposal for cosmology has been taken very seriously and over the last fifteen years $f(R)$ gravity has been one of the most active areas of research in cosmology.

Now it can easily be shown that $f(R)$ theory is equivalent to
scalar tensor theory, or more precisely BD theory
with vanishing BD coupling parameter $\omega = 0$. This would seem to be a problem for the theory because when
$\omega=0$, the post-Newtonian parameter (PPN) $\gamma_{BD} = \frac{1 + \omega}{2 + \omega} = 1/2$ and this is very different from  the value $\gamma\sim1$ required by Solar System tests. However it must be pointed out \cite{olmo06,faraoni08} that the $f(R)$ theory is not equivalent to the standard
BD-theory with vanishing scalar potential. In fact the PPN parameter for $f(R)$ theory is given by a more complicated expression than that of the BD theory \cite{capozziello10}. In this case $\omega=0$ is accompanied by a non-zero $V(\phi)$ and
so depending on the mass of the scalar field which is included in $V(\phi)$, the theory can  agree with the Solar System tests. Hence introducing the scalar field and associated potential by
\begin{eqnarray}
\phi & \equiv & f_R(R) \nonumber \\
V(\phi) & \equiv & R(\phi)f_R - f(R(\phi)) \label{Vphi}
\end{eqnarray}
the field equations (\ref{fReqs}) and (\ref{trace}) become
\begin{eqnarray}
R_{\mu\nu} - \frac{1}{2}g_{\mu\nu}R & = & \frac{\kappa}{\phi}T_{\mu\nu} - \frac{1}{2\phi}g_{\mu\nu}V(\phi) \nonumber \\
& & + \frac{1}{\phi}[\nabla_{\mu}\nabla_{\nu}\phi - g_{\mu\nu}\Box\phi], \label{BDeqs}
\end{eqnarray}
and
\begin{equation}
3\Box\phi + 2V(\phi) - \phi\frac{dV}{d\phi} = \kappa T \label{BDeqs2}
\end{equation}
respectively, where $dV/d\phi = R$. The above equations are recognized as the field equations of BD theory with $\omega = 0$ with a non-trivial potential, whose action takes the form
\begin{equation}
S = \frac{1}{2\kappa}\int d^4x\sqrt{-g}[\phi R - V(\phi)] + S_m.
\end{equation}
In order to obtain the scalar potential $V(\phi)$ the equation $\phi = f_R(R)$ needs to be invertible, and conversely a given BD theory having $\omega=0$ can be expressed as an $f(R)$ theory provided that the equation $dV(\phi)/d\phi = R$ is invertible.

\section{No hair theorem for $R^2$ black holes}
As described in the Introduction, the proof follows the lines of a recent no hair result by Faraoni \cite{faraoni17}  for static and spherically symmetric black holes in the generalized Brans-Dicke theory with a variable coupling term $\omega(\phi)$. However in that case, in order to integrate the field equations, the author assumed a zero scalar potential $V(\phi)$. In our case $\omega = 0$ but as seen from Eq. (\ref{Vphi}), $V(\phi)$ is non-zero for $f(R) = R^2$. So we start with a general static and spherically symmetric line element
\begin{equation}
ds^2 = -A^2(r)dt^2 + B^2(r)dr^2 + r^2(d\theta^2 + \sin^2(\theta)d\phi^2).
\end{equation}
In this case the position of the apparent horizon (which coincides with the event horizon due to the static geometry) is given by the locus of points satisfying $\nabla_\alpha r\nabla^\alpha r =0$, i.e. $g^{rr}=0$ or $B\rightarrow\infty$. To simplify the field equations we consider
a source with a trace-free energy momentum tensor such that $T=0$. This can be represented by an electromagnetic field or a massless conformally coupled scalar field $\psi$ having a zero or quartic potential. According to (\ref{Vphi}) when $f(R) = R^2$, $\phi = 2R$ and so $\phi(r)$ depends only on the radial coordinate\footnote{Note that the fact that the metric is static and spherically symmetric does not imply that the scalar field is static as well. In fact there are various scalar field solutions in scalar tensor gravity with a time dependent scalar field. See for example \cite{wyman81,beato06,robinson06,maeda12}.} $r$. The field equation (\ref{BDeqs2}), then takes the form
\begin{equation}
3\left[\phi'' + \left(\frac{A'}{A} - \frac{B'}{B} + \frac{2}{r}\right)\phi'\right] = B^2(\phi \frac{dV}{d\phi} - 2V). \label{geneq}
\end{equation}
From (\ref{Vphi}) the scalar field potential is $V(\phi) = \phi^2/4$ and so the RHS of the above field equation vanishes. Hence it takes the simple form
\begin{equation}
\left[\ln\left(\frac{Ar^2\phi'}{B}\right)\right]'=0,
\end{equation}
which can be integrated to give
\begin{equation}
\frac{\phi'(r)}{B} = \frac{C}{Ar^2},
\end{equation}
where $C$ is a constant of integration. Now on the event horizon $1/B \rightarrow 0$ and therefore if $\phi$ and $\phi'$ are finite there, then the above relation implies that $C=0$ and so $\phi'(r)$ is zero everywhere outside the event horizon, meaning that the scalar field takes a constant value. In this case the BD field equations (\ref{BDeqs}) reduce to Einstein's field equations with a cosmological term and so the black hole becomes the Schwarzschild (anti)de-Sitter black hole or the Reissner-Nordst\"{o}m (anti)de-Sitter black hole.

\section{A black hole solution in $R^2$ gravity with a conformally coupled scalar field}
In the presence of a non-minimally coupled scalar field $\psi$ the general action is given by (taking $\kappa = 8\pi G = c =1$)
\begin{equation}
S = \int\mathrm{d}^4x\sqrt{-g} \frac{f(R,\psi)}{2} - \frac{1}{2}\nabla_{\alpha}\psi\nabla^{\alpha}\psi - V(\psi) + S^{(m)}, \label{genaction}
\end{equation}
where $f(R,\psi)$ is a function of the Ricci scalar $R$ and a nonminimally coupled scalar field $\psi$. Sometimes this is referred to as the \textit{generalized scalar tensor theory}. The field equations obtained by varying the action with respect to the metric $g_{\mu\nu}$ and the scalar field $\psi$ are given respectively by \cite{mathew17}
\begin{eqnarray}
\lefteqn{f_RG_{\mu\nu} = \nabla_{\mu}\psi\nabla_{\nu}\psi - \frac{1}{2}g_{\mu\nu}\nabla^{\alpha}\nabla_{\alpha}\psi}\nonumber \\
& & +\frac{1}{2}g_{\mu\nu}(f - f_R R - 2V(\psi)) + \nabla_{\mu}\nabla_{\nu}f_R - g_{\mu\nu}\Box f_R, \label{Frpsi}
\end{eqnarray}
and
\begin{equation}
\Box\psi + \frac{1}{2}\left(\frac{\partial f}{\partial \psi} - 2\frac{dV}{d\psi}\right)=0,
\end{equation}
where the subscript in $f_R$ denotes partial derivative with respect to $R$. As in the minimally coupled case the effective gravitational coupling will be $G_{\mathrm{eff}} = G/f_R$ such that (\ref{Frpsi}) takes the Einstein-form (\ref{einstein-form}) with $T^{\mathrm{eff}}_{\mu\nu}$ given by the same expression in (\ref{emcfluid}) with $f(R)$ replaced by $f(R,\psi)$.

Now omitting ordinary matter and taking a conformally coupled scalar field in pure $R^2$ gravity such that $f(R,\psi) = R^2 + \xi R\psi^2$; $\xi=1/6$ in (\ref{genaction}), the above field equations become
\begin{eqnarray}
\lefteqn{2G_{\alpha\beta}R + \frac{1}{2}g_{\alpha\beta}R^2 - 2R_{;\alpha\beta} + 2g_{\alpha\beta}\Box R - \psi_{\alpha}\psi_{\beta} + \frac{1}{2}g_{\alpha\beta}\nabla^{\gamma}\psi\nabla_{\gamma}\psi}\nonumber \\
 & &  + g_{\alpha\beta}V(\psi) - \frac{1}{6}(g_{\alpha\beta}\Box\psi^2 - \psi^2_{;\beta\alpha} + G_{\alpha\beta}\psi^2) =0 \label{R2eqs}
\end{eqnarray}
and
\begin{equation}
\Box\psi - \frac{1}{6}R\psi - \frac{dV}{d\psi} = 0, \label{R2scl}
\end{equation}
respectively.
For a quartic potential $V(\psi) = \lambda\psi^4$, the above field equations are solved by the following static and spherically symmetric line element
\begin{equation}
ds^2 = -\left[-\frac{\Lambda}{3}r^2 + \left(1 - \frac{M}{r}\right)^2\right]dt^2 + \left[-\frac{\Lambda}{3}r^2 + \left(1 - \frac{M}{r}\right)^2\right]^{-1}dr^2 + r^2(d\theta^2 + \sin^2\theta d\phi^2) \label{rn}
\end{equation}
with the scalar field given by
\begin{equation}
\psi = \frac{4\sqrt{3\Lambda}M}{r - M} \label{rnpsi}
\end{equation}
and $\lambda = -1/288$.
The line element coincides with the extreme case ($Q=M$) of the Reissner-Nordstr\"{o}m-de Sitter black hole solution in GR with the inner and outer event horizons at
\begin{equation}
r_{\pm} = \frac{l}{2}(\pm 1 \mp \sqrt{1 \mp 4M/l})
\end{equation}
and a cosmological horizon at $r_c = \frac{l}{2}(1 + \sqrt{1 - 4M/l})$, where $l = \sqrt{3/\Lambda}$. The scalar field is regular everywhere outside the event horizon and is singular at $r = M$ which is located between the two event horizons. The metric has a curvature singularity at $r=0$. This metric has also been shown \cite{martinez03} to be a solution of the field equations in GR with a conformally coupled scalar field $\phi = \sqrt{3/4\pi}\sqrt{G}M/(r - GM)$ having a quartic potential $V(\phi) = \alpha\phi^4$, where $\alpha = -\frac{2}{9}\pi\Lambda G$. For the metric in (\ref{rn}) the Ricci scalar $R$ is constant ($R=4\Lambda$), and so in this case the field equations in (\ref{R2eqs}) and (\ref{R2scl}) reduce to those of GR with a cosmological constant and a conformally coupled scalar field. So one would expect that the solution in (\ref{rn}) and (\ref{rnpsi}) is the same solution obtained in Ref.\cite{martinez03}. Indeed redefining $\phi = \psi/\sqrt{8\Lambda}$ and $\alpha = 8\Lambda\lambda$ and using the fact that in our case we let $8\pi G=1$, it is easy to show that the two solutions are equivalent. However this doesn't happen in the case when the Ricci scalar $R=0$, i.e., $\Lambda=0$. In the GR case the solution in Ref. \cite{martinez03} reduces to the well know BMBB solution found by Bronnikov, Melnikov and Bocharova \cite{bocharova70} and independently by Bekenstein \cite{bekenstein74}. In our case the BBMB solution is not a solution of pure $R^2$ gravity. It should be noted that the scalar field in (\ref{rnpsi}) is not defined in terms of an extra global ``charge'' (the constants of integration $M$ and $\Lambda$ appear in the metric) and so it does not endow the black hole with hair. Letting $M\rightarrow0$ or $\Lambda\rightarrow0$ results in a vanishing scalar field. The first limit leads to the (A)dS metric as a solution of $R^2$ gravity with a constant Ricci scalar, which as noted above is equivalent to GR. The second limit leads to the extreme Reissner-Nordstr\"{o}m metric, again as a solution of $R^2$ gravity, but now with a vanishing Ricci scalar.

\section{Conclusion}
In this paper we have proved a NHT for the scale-invariant ghost free pure $R^2$ gravity. This extends a recent result by Ca\~{n}ate \cite{canate17} who recently extended the no-hair result to $f(R)$ theory, which however excluded the case $f(R)=R^2$. Moreover unlike previous NHTs in GR and other alternative theories including Ca\~{n}ate's, our proof does not place any restrictions on the black hole asymptotics. When proving NHTs, it is usually very difficult to avoid the assumption of asymptotic flatness, and even de Sitter asymptotics may be problematic. Doing away with these restrictions would be desirable for the future study of more realistic black holes embedded in FLRW backgrounds. In order to be able to integrate the field equations we limited our result to static and spherically symmetric black holes. We used the equivalence between general $f(R)$ theory and scalar tensor theories. The result is also based on a proof by Faraoni \cite{faraoni17} of the NHT for spherical scalar-tensor black holes with vanishing scalar potential. We have also presented an example of a static spherically symmetric black hole in $R^2$ gravity having a conformally coupled scalar field with associated Higgs-like potential. The metric in this case is identical to the extreme Reissner-Nordstr\"{o}m-de Sitter solution and the scalar field is regular everywhere outside the horizon and vanishes asymptotically. This solution is also equivalent to that obtained in Ref. \cite{martinez03} in the case of GR with a cosmological constant and a conformally coupled scalar field. The conformally coupled scalar field would not qualify as an example of hair, considering that it is regulated by the same parameters that appear in the metric.

\section*{Acknowledgments}
J.S. gratefully acknowledges financial support from the University of Malta during his visit at NASA-GSFC and the hospitality of the Astrophysics Science Division of GSFC.

\end{document}